\documentstyle{article}
\font\elvmathbold=cmmib10 scaled\magstephalf
\def\bbeta{\mbox{\elvmathbold\char'014}}
\def\ggamma{\mbox{\elvmathbold\char'015}}

\def\setfonts{%
\font\frbig=eufm10 scaled\magstep1
\font\frscr=eufm10
\font\frscrscr=eufm8
\newfam\frfam
\textfont\frfam=\frbig
\scriptfont\frfam=\frscr
\scriptscriptfont\frfam=\frscrscr
\def\fr{\fam\frfam}
\font\openbig=msbm10 scaled\magstep1
\font\openscr=msbm10
\font\openscrscr=msbm8
\newfam\openfam
\textfont\openfam=\openbig
\scriptfont\openfam=\openscr
\scriptscriptfont\openfam=\openscrscr
\def\open{\fam\openfam}
}

\font\tenrm  = cmr10
\font\tenmi  = cmmi10
  \skewchar\tenmi ='177
\font\tensy  = cmsy10
  \skewchar\tensy ='60
\font\tenex  = cmex10

\font\tensf  = cmss10
\font\tenly  = lasy10

\font\egtsf=cmss8

\font\sevrm  = cmr7
\font\sevmi  = cmmi7
  \skewchar\sevmi ='177
\font\sevsy  = cmsy7
  \skewchar\sevsy ='60

\font\fivrm  = cmr5
\font\fivmi  = cmmi5
  \skewchar\fivmi ='177
\font\fivsy  = cmsy5
  \skewchar\fivsy ='60

\setfonts

\makeatletter

\def\tableofcontents{\subsection*{\mbox{}
} \@starttoc{toc}}
\newdimen\normalarrayskip
\newdimen\minarrayskip
\normalarrayskip\baselineskip
\minarrayskip\jot
\newif\ifold \oldtrue \def\new{\oldfalse}
\def\arraymode{\ifold\relax\else\displaystyle\fi}

\def\@arrayskip{\ifold\baselineskip\z@\lineskip\z@
  \else
  \baselineskip\minarrayskip\lineskip2\minarrayskip\fi}
\def\@arrayclassz{\ifcase \@lastchclass \@acolampacol \or
\@ampacol \or \or \or \@addamp \or
 \@acolampacol \or \@firstampfalse \@acol \fi
\edef\@preamble{\@preamble
 \ifcase \@chnum
  \hfil$\relax\arraymode\@sharp$\hfil
  \or $\relax\arraymode\@sharp$\hfil
  \or \hfil$\relax\arraymode\@sharp$\fi}}
\def\@array[#1]#2{\setbox\@arstrutbox=\hbox{\vrule
  height\arraystretch \ht\strutbox
  depth\arraystretch \dp\strutbox
  width\z@}\@mkpream{#2}\edef\@preamble{\halign \noexpand\@halignto
\bgroup \tabskip\z@ \@arstrut \@preamble \tabskip\z@ \cr}%
\let\@startpbox\@@startpbox \let\@endpbox\@@endpbox
 \if #1t\vtop \else \if#1b\vbox \else \vcenter \fi\fi
 \bgroup \let\par\relax
 \let\@sharp##\let\protect\relax
 \@arrayskip\@preamble}
\def\l@section#1#2{\addpenalty{\@secpenalty} \addvspace{.4em plus 1pt}
\@tempdima 1.5em \begingroup
 \parindent \z@ \rightskip \@pnumwidth
 \parfillskip -\@pnumwidth
 \small\sf \leavevmode \advance\leftskip\@tempdima
 \hskip -\leftskip #1\nobreak\hfil
\nobreak\hbox to\@pnumwidth{\hss #2}\par
 \endgroup}
\def\tenpt{%
\textfont\z@\tenrm
  \scriptfont\z@\sevrm \scriptscriptfont\z@\fivrm
\textfont\@ne\tenmi \scriptfont\@ne\sevmi \scriptscriptfont\@ne\fivmi
\textfont\tw@\tensy \scriptfont\tw@\sevsy \scriptscriptfont\tw@\fivsy
\textfont\thr@@\tenex \scriptfont\thr@@\tenex \scriptscriptfont\thr@@\tenex
\def\unboldmath{\everymath{}\everydisplay{}\@nomath\unboldmath
  \textfont\@ne\tenmi
  \textfont\tw@\tensy \textfont\lyfam\tenly
  \@boldfalse}\@boldfalse
\def\boldmath{\@ifundefined{tenmib}{\global\font\tenmib\@mbi
   \global\font\tensyb\@mbsy
   \global\font\tenlyb\@lasyb\relax\@addfontinfo\@xpt
   {\def\boldmath{\everymath{\mit}\everydisplay{\mit}\@prtct\@nomathbold
  \textfont\@ne\tenmib \textfont\tw@\tensyb
  \textfont\lyfam\tenlyb \@prtct\@boldtrue}}}{}\@xpt\boldmath}%
\def\psf{\fam\sffam\tensf}\textfont\sffam\tensf
  \scriptfont\sffam\egtsf \scriptscriptfont\sffam\egtsf
}
\makeatother

\def\lvm{\leavevmode\hbox to\parindent{\hfill}}
\def\req#1{(\ref{#1})}

\def\BE{\begin{equation}}
\def\EE{\end{equation} }
\def\BA{\begin{array}}
\def\EA{\end{array}}
\def\BB{\begin{equation}\new\begin{array}{rcl}}
\def\L{\left}
\def\R{\right}

\def\frac#1#2{{\textstyle{{#1}\over{#2}}}}

\def\ket#1{\bigl|{#1}\bigr\rangle}

\def\d{\partial}

\def\half{{\textstyle{1\over2}}}

\def\fourth{{\textstyle{1\over4}}}

\def\threehalves{{\textstyle{3\over2}}}

\def\cF{{\cal F}}
\def\cG{{\cal G}}
\def\cK{{\cal H}}
\def\cH{{\cal H}}
\def\cJ{{\cal J}}
\def\cL{{\cal L}}

\def\cQ{{\cal Q}}

\def\cT{{\cal T}}

\def\cX{{\cal X}}

\def\oT{{\open T}}

\def\cTvee{\cT}
\def\cGvee{\cG^0}

\def\htop{{\sf h}}

\def\cc{central charge}

\def\emt{energy-momentum tensor}

\def\up#1{^{\rm[#1]}}
\def\ap{\alpha_+}
\def\am{\alpha_-}
\def\an{\alpha_0}

\def\tensor{\otimes}
\def\vst{v_\ast}

\def\Sst{{\cal S}_\ast}
\def\st{\sqrt{2}}
\def\tilde{\widetilde}

\def\NPB{Nucl. Phys. B}
\def\PLB{Phys. Lett. B}
\def\MPLA{Mod. Phys. Lett. A}
\def\CMP{Commun. Math. Phys.}
\def\IJMPA{Int. J. Mod. Phys. A}
\def\MPLA{Mod. Phys. Lett. A}

\def\noi{\noindent}

\begin{document}
\hfuzz=1.8pt

\begin{flushright}
{\tt hep-th@xxx}/9410109
\end{flushright}
\begin{center}
{\Large\sc Inverting the Hamitonian Reduction\\[5pt]
in String Theory}\footnote{
Talk given at the $28^{\rm th}$ International Symposium on the
Theory of Elementary Particles,
Wendisch-Rietz, August 30 -- September 3, 1994}\\[2ex]
\large A.~M.~Semikhatov\\[0.5ex]
{\small\sl I.E.~Tamm Theory Division, P.~N.~Lebedev Physics Institute\\
Russian Academy of Sciences, 53 Leninski prosp., Moscow 117924, Russia}
\\[2ex]
\parbox{.9\hsize}{\addtolength\baselineskip{-2ex}
\noindent\small It is well known that many interesting realisations of
string
theories can be obtained via hamiltonian reduction from WZW models.
I want to
point out that string theories do in certain cases also provide the recipe
to reconstruct the ambient space of the hamiltonian reduction,
including Ka\v{c}--Moody currents  and the associated ghosts.  The
procedure of reconstructing the Ka\v{c}--Moody currents is closely related
to
properties of matter+gravity multiplets in noncritical string theories. In
application to KPZ gravity and its $N\!=\!1$ supersymmetric extension, the
`inverted hamiltonian reduction' constructions serve to establish relation
with the DDK-type formalism for matter + gravity.  }
\end{center}

\thispagestyle{empty}
\subsection*{Introduction}\lvm
The importance of embedding of non-critical strings into
WZW models was pointed out recently in~\cite{[BLLS]}.
Such embeddings would in a number of cases look like
inverting the quantum Hamiltonian
reduction~\cite{[BO],[dBT],[BLNW],[L-rings]}.
Generally, `inverting the Hamiltonian reduction' is an ill-posed problem,
 as is the inversion of any projection:
\BE\new\BA{ccc}
K&\longrightarrow&G\\
{}&{}&\biggm\downarrow\lefteqn{\pi}\\
{}&{}&G/K
\EA\EE
Constructing $\pi^{-1}$ involves non-canonical choices of representatives
etc. However, in conformal field theory an
`almost canonical' choice for these non-canonical data
happens to be provided by coupling to (super)gravity.
Matter + gravity theories possess an `$N\mapsto N+2$ supersymmetry
enhancement'~\cite{[GS3],[BLNW]}.
While not every $N+2$-supersymmetric model is necessarily an
$N$-matter dressed with gravity, {\it demanding\/} it to be so
provides one with almost all the data necessary to invert the hamiltonian
reduction. The only remaining piece is then provided by
tensoring the theory with a Fock module that is a
`bosonisation' of the matter Verma module.
In this way the ambient space of the
reduction, i.e. the original Ka\v{c}--Moody algebra {\it and\/} the
`Drinfeld--Sokolov' ghosts that are necessary to build up the BRST
 complex,
can be reconstructed,
at least for those cases when the
hamiltonian reduction results in a linear algebra.

As a `practical' application of the `inverted Hamiltonian reduction'
I will consider, following \cite{[HS]}, the KPZ
formulation \cite{[P-gr],[KPZ],[DJNW],[Itoh],[Ku],[Horv],[PZ],[Rsh]}
of the induced (super)gravity.
While the above-mentioned
`supersymmetry enhancement' can be considered as a fundamental
property of gravity in the conformal gauge \cite{[Da],[DK]},
another fundamental property of two-dimensional
gravity is the existence, in the light-cone gauge,
of an $s\ell(2)$ Ka\v{c}--Moody symmetry or its
super-extension \cite{[PZ],[Rsh]}.
While the $s\ell(2)$ current algebra, as it was derived,
appears completely independent of the $N\!=\!2$ formalism,
one may nevertheless wonder what is the relation
between conformal field-theoretic descriptions of the two formalisms for
two-dimensional gravity. They cannot be literally embedded into one
another,
as the underlying symmetry algebras would not allow that.

Recently, a more
general concept of relations between conformal field theories
has emerged in
the context of Universal String Theory
\cite{[BV],[Fof],[IO],[OP],[BOP1],[BFW],[H-new],[BerkOh]}.
In this approach, one realises
theories with lower symmetries as `special backgrounds' of those with
higher
symmetries.
This construction actually considers conformal field theories
modulo trivial topological theories: the extra fields that
`decouple' do in fact form a trivial topological sector, which does not
bring in any new physical states.
In the spirit of Universal String Theory applied to the
{\it gravity\/} sector, I will `evaluate' the KPZ theory on
a background provided by a
representation for the current algebra
obtained by the `inverted Hamiltonian reduction'.
The result turns out to be the DDK formulation of matter+gravity
models, plus an extra topological piece. Thus the DDK theory arises as a
`special background' of the KPZ theory.  Conversely,
when the (super)DDK theory is tensored with a certain topological field
theory, a
hidden (super-)$s\ell(2)$ symmetry emerges.  It will be discussed
below in which
sense the `extra' topological theory can be considered {\sl trivial\/};
this
involves a fermionic screening operator similar to the one discussed in
\cite{[BLNW],[L-rings],[EKYY]}.

\smallskip

This talk is based on a work with Chris Hull \cite{[HS]}.

\subsection*{Inverting the hamiltonian reduction} \lvm
$\bullet$~The {\bf topological conformal algebra} \cite{[Ey],[W-top]}
is generated by an
\emt\ $\cT$, a bosonic dimension-one current $\cH$, a fermionic
dimension-one
(BRST)
current $\cJ$ and a fermionic dimension-two current $\cG$.
These generators are realised in any
conformal field theory made up by a matter theory with \emt\
$T$ of central charge~$d$, $b c $ ghosts that
have dimensions $1,0$ \cite{[GS3]},
and a single scalar $\phi$ with background charge $\alpha_0$
chosen so that the total central charge vanishes, $d-2+(1+6\alpha_0^2)=0$.

The generators of the topological conformal algebra read
\BE \new\BA{rcl}
\cT&=&{}T-\half(\d\phi)^2+{\an\over\st}\d^2\phi-b \d c \,,\qquad
\cJ~{}={}~b \,,\\
\cG&=&{}c \Bigl(T-\half(\d\phi)^2+{\an\over\st}\d^2\phi\Bigr)
-b \d
c \!\cdot\!c  +\st\ap\d
c \!\cdot\!\d\phi+\half(1-2\ap^2)\d^2c \,,\\
\cH&=&{}-b c -\st\ap\d\phi \EA\label{spin1} \EE
The
parameters $\alpha_{\pm,0}$ satisfy $\an=\ap+\am $, $\ap\am=-1$ as usual,
and
it will be useful to introduce a level $k$ by
$\ap={1\over\sqrt{k+2}}$, $\am=-\sqrt{k+2}$.

\medskip

$\bullet$~I am going to show that by adding an extra scalar $\vst$ to the
system
of fields from the RHSs or \req{spin1}, it is
possible to {\bf construct an $s\ell(2)_k$ algebra}, and in fact a
(twisted)
${sl}(2)_k\oplus u(1)_{BC}$ algebra (where $u(1)_{BC}$ is actually a
bosonisation
of a pair of fermionic ghosts).

Among the building blocks there are, therefore, an \emt\ $T$ with central
charge $d=1-6\an^2$, a `Liouville' scalar $\phi$ with background charge
$\an/ \sqrt 2$, a pair of ghosts $b ,c $ of
dimensions $1,0$, and the $\vst$ scalar.
Then a twisted $N\!=\!2$ conformal algebra is generated by
\req{spin1}, while the
${sl}(2)_k$ currents are constructed as
\BE\new\BA{rcl}
J^+&=&e^{\sqrt{2}\ap(\vst-\phi)}\,,\qquad
J^0~{}={}~b c +\sqrt{2}\ap\d\phi
-\biggl({\am\over\st}+\st\ap\biggr)\d\vst\,,\\
J^-&=&\biggl\{\am^2\Bigl(T-\half(\d\phi)^2+{\an\over\st}\d^2\phi\Bigr)
+2b \d
c -\am^2\d(b c ) +\st\am\d\phi\!\cdot\!b c \biggr\}
e^{-\sqrt{2}\ap(\vst-\phi)}
\EA\label{sl2}\EE
Moreover, the `Drinfeld--Sokolov' $BC$ ghosts (which are a fermionisation
of the
$u(1)$ current), in turn, can be
constructed as \BE B=c e^{\sqrt{2}\ap(\vst-\phi)}\,,\qquad
C=b e^{-\sqrt{2}\ap(\vst-\phi)}.\label{BC} \EE

The formulae \req{sl2}--\req{BC} invert the (quantum) Drinfeld--Sokolov
reduction, in the sense that, given an abstract matter theory that
represents
the result of the reduction ($J^-\sim T$), eqs.~\req{sl2}, \req{BC} show
how to add `dressing' fields that would lead to reconstructing the ambient
space of the hamiltonian reduction, together with the ghost system that is
required in order to build up the Drinfeld--Sokolov BRST complex.

Evaluating the twisted Sugawara \emt\ \BE\widetilde{T^{{\rm{S}}}}={1\over
k+2}\L(J^0J^0+\half(J^+J^-+J^-J^+)\R)+\d J^0\,.\label{Tsug}\EE with the
currents being given by \req{sl2}, one arrives at the identity \BE
\tilde{T^{\rm S}}+\d B\!\cdot\!C=\cT+\half(\d\vst)^2-{\an\over\st}\d^2\vst
\label{Tidentity} \EE
where $\cT$ is the \emt\ from~\req{spin1}.  Eq.~\req{Tidentity} thus
represents
two ways to describe the same theory: $s\ell(2)\oplus
u(1)_{BC}\simeq {\rm(topological)}\oplus(\d\vst)$.

\smallskip

Moreover, admissible $s\ell(2)$ representations can be
arrived at starting from topological algebra representations tensored
with the
$\vst$-matter sector.
BRST-invariant primary states of the topological conformal algebra
are characterised by their topological $U(1)$ charge:
\BE
\cH_0\ket\htop_{\rm
top}=\htop\ket\htop_{\rm top}\,,\qquad \cL_{\geq0}\ket\htop_{\rm
top}=\cH_{\geq1}\ket\htop_{\rm top}=\cG_{\geq1}\ket\htop_{\rm
top}=\cJ_{\geq0}\ket\htop_{\rm top}=0\,.
\label{cpsconditions}
\EE
In the representation
\req{spin1} of the topological algebra,
such  states are
constructed from matter dressed with ghosts and Liouville, and are
given by
\BE \ket{\htop(r,s)}_{\rm top}= \ket{r,s}\tensor\ket{p_{\rm M}(r,s)}_{\rm
L}\tensor\ket0_{bc}\,, \label{keth} \EE
where $p_{\rm M}(r,s)=-{1\over\st}(\ap(r-1)+\am(s-1))$,
and $\ket{p}_{\rm L}$ corresponds to $e^{p\phi}$ in the Liouville sector.
The topological $U(1)$ charge is evaluated as
$\htop(r,s)=-\ap^2(r-1)+s-1$.
Admissible $s\ell(2)_k$
highest-weight primary states of spin \BE
j(r,s)={r-1\over2}-(k+2){s-1\over2}\label{jtrue} \EE are arrived at
(tensored with the ghost vacuum $\ket0\equiv\ket0_{BC}$) as follows:
\BE
\ket{j(r,s)}_{s\ell(2)}\tensor\ket0_{BC}=
\ket{r,s}\tensor\ket{p_{\rm M}(r,s)}_{\rm
L}\tensor\ket0_{bc} \tensor\ket{-p_{\rm M}(r,s)}_\ast\label{ketket} \EE
On the LHS of \req{ketket}, we have thus
obtained a primary state of the algebra $s\ell(2)_k\oplus u(1)_{BC}$.

The RHS of \req{ketket} can also be read as a dressing of the matter
operator
$U_{r,s}\sim\ket{r,s}$ with \BE V_{r,s}=e^{-p_{\rm M}(r,s)(\vst-\phi)}=
e^{j(r,s)\sqrt{{2\over k+2}}(\vst-\phi)}\,.\label{vphi}\EE


$\bullet$~To extend the previous constructions to {\bf $N\!=\!1$
supergravity},
I start with an
$N\!=\!1$ matter that comprises an \emt\ $T_{\rm m}$ and its superpartner
$G_{\rm m}$:
\BE\new\BA{rcl} T_{\rm m}(z) T_{\rm m}(w) &=& {d/2 \over(z-w)^4} +
{2 T_{\rm
m}(w)\over(z-w)^2} + {\d T_{\rm m}(w)\over z-w}\,,\\ T_{\rm m}(z) G_{\rm
    m}(w) &=& {3/2\,G_{\rm m}(w)\over(z-w)^2} +
{\d G_{\rm m}(w)\over z-w}
    \,,\qquad
G_{\rm m}(z) G_{\rm m}(w)~{}={}~{2d/3 \over(z-w)^3} + {2 T_{\rm
m}(w)\over z-w}\,, \EA\label{matter}\EE
where $d = {15\over2} - 3\am^2 - 3\ap^2$
is the matter cental charge and $\am=-1/\ap$.
When coupling this system to supergravity, one introduces a super-Liouville
field, with components $\phi$, $\psi$,
and fermionic
and bosonic ghosts $bc$ and $\beta\gamma$.
The \emt\ is given by \BE
\cT=T_{\rm m} - \half\d\phi\,\d\phi  + \half(\ap-\am)\d^2\phi -
   \half\d\psi\,\psi - \d b\,c - 2b\d c - \threehalves\beta \d \gamma-
   \half\d\beta\,\gamma\,.\label{Tl}\EE

The full theory admits an $N\!=\!3$ symmetry algebra whose generators can be
constructed as explained in \cite{[BLNW]}.  First, the three supersymmetry
generators are given by
\BE\new\BA{rcl} \cG^+&=&b\,,\qquad\cG^0~{}={}~{}-G_{\rm m} +
  b\gamma - 2c\d \beta  - 3\d c\beta - \psi\d\phi + \left( \ap-\am \right)\d
   \psi \,,\\ \cG^-&=& 4cT_{\rm m}+ 2\gamma G_{\rm m} -b\gamma \gamma  -
4b\d
  c\,c - 2c\beta\d\gamma  - 2c\d\phi\,\d\phi  + 2c\d\beta\,\gamma  -
   2c\d\psi\,\psi   + (-2\am + 2\ap - 8x) c\d^2\phi \\ {}&{}&{}  +
   2\psi\gamma\d\phi + 4x\psi\d\gamma   - 8x\d c\,\d\phi  + 4\d
   c\,\beta\gamma  + ( 2\am - 2\ap + 4x)\d\psi\,\gamma  -
   8x^2\d^2c \EA\label{GGG}\EE where two values of the parameter $x$ are
possible, $x=\half\ap$ or $x = -\half\am$.
Next, there exists a Majorana--Weyl fermion
$\cF =
-2x\psi + c\beta$.
Operator products
of the supersymmetry generators with $\cF$ generate three bosonic
currents,
$\cG^a(z)\cF(w) = {\cK^a\over z-w}$, where
\BE\new\BA{rcl}
\cK^+&=&\beta \,,\qquad \cK^0~{}={}~{}-2x\d\phi
+ bc + \beta \gamma\,,\\ \cK^-&=&
  -2bc\gamma  - \beta \gamma \gamma  - 2c\psi \d\phi  + ( -2\am + 2\ap -
8x
   )c\d \psi  + 4x\gamma\d\phi  + 2G_{\rm m}c - 4x\d c\,\psi  + 2\d
   c\,c\beta
+ 8{x^2}\d\gamma\,.\EA\label{KKK}\EE These satisfy an
$s\ell(2)$ algebra at level $-8x^2$:
\BE\cK^0(z)
\cK^\pm(w)=\pm{\cK^\pm\over z-w}\,,\quad
\cK^0(z)\cK^0(w)={-4x^2\over(z-w)^2}\,,\quad \cK^+(z)\cK^-(w)=
{-8x^2\over(z-w)^2}+{2 \cK^0\over z-w}
\label{KK}\EE

Now, before completing the list of commutation relations among the
$N\!=\!3$
generators, observe that the (twisted, to be precise) $N\!=\!3$ algebra
admits an involutive automorphism that consists in
$\cG^\pm\mapsto\cG^\mp$,
$\cK^\pm\mapsto\cK^\mp$, $\cK^0\mapsto-\cK^0$, and twisting
$\cT\mapsto\cT+\d \cK^0$, $\cG^0\mapsto-\cG^0 - 2\d \cF$.
As I will need only the second,
transformed,  version, I will not distinguishing between the two
constructions related by the automorphism, and will simply consider the
above
constructions for $\cG^\pm$ and $\cK^{\pm,0}$ as belonging to the algebra
that contains the twisted \emt\ \BE\cTvee= T_{\rm m}  - \half \d\phi
\d\phi +
\half(\ap - \am - 4x) \d^2\phi - \half\d \psi\,\psi - b\d c -
\half\beta \d
\gamma   + \half\d \beta\,\gamma \label{Ttw}\EE and the $\cG^0$
generator
\BE\cGvee = -G_{\rm m} + b\gamma - \d c\beta - \psi \d\phi  + (
   \ap-\am - 4x)\d\psi \EE (the previous $\cT$ and $\cG^0$ will not appear
any more).  I also choose the first option
for the parameter from \req{GGG}
$x=\half\ap$.  It can be checked then that, in addition to
\req{KK}, the following relations hold:  \BE\new\BA{rclcrcl}
\cK^+(z)\cGvee(w)&=&{-\cG^+\over z-w}\,,&{}&
 \cK^+(z)\cG^-(w)&=&{2\cF(w)\over(z-w)^2} + {2\cGvee-2\d \cF\over z-w}
\,,\\
\cK^0(z)\cG^\pm(w)&=&{\pm \cG^\pm\over z-w}\,,&{}&
 \cK^0(z)\cGvee(w)&=&{\cF(w)\over(z-w)^2}\,,\\
\cK^-(z)\cG^+(w)&=&{2\cF(w)\over(z-w)^2} + {-2\cGvee + 2 \d \cF\over z-w}
\,,
&{}&\cK^-(z)\cGvee(w)&=&{\cG^- \over z-w}\,.  \EA\EE and
\BE\new\BA{rclcrcl}
\cGvee(z)\cG^+(w)&=&{\cK^+(w)\over(z-w)^2}+{\d \cK^+\over z-w}\,,
&{}&\cG^+(z)\cG^-(w)&=&{-16x^2\over(z-w)^3}+{4\cK^0(w)\over(z-w)^2}
+{4\cTvee\over z-w}\,,\\ \cGvee(z)\cG^-(w)&=&{-3\cK^-(w)\over(z-w)^2} +
{-\d
\cK^-\over z-w}\,, &{}&\cGvee(z)\cGvee(w)&=&{2\cTvee\over z-w}\,.
\EA\EE The
\emt~\req{Ttw} assigns $(\cG^+, \cGvee, \cG^-; \cK^+, \cK^0, \cK^-)$
dimensions $(1, \threehalves, 2;\half, 1, \frac{3}{2})$.

\smallskip

$\bullet$~Now I proceed to the {\bf construction of an $osp(1|2)$ algebra}
and the associated `Drinfeld--Sokolov' ghosts.
Just as in the
purely bosonic case, let us add to the system of supermatter coupled to
$N\!=\!1$ supergravity a matter theory explicitly represented by a scalar
field $\d\vst$
with the \emt\
\BE T_\ast=\half \d \vst\d \vst  + \half(\ap + \am)\d^2\vst \EE
Then, the $osp(1|2)$ currents are constructed as \BE\new\BA{rcl}
J^+&=&e^{2\ap(\phi-\vst)}\,,\quad
j^+~{}={}~{\sqrt{2}}\psi e^{\ap(\phi-\vst)}\,,\quad
J^0~{}={}~\half\am\d\phi + bc +\half\am{\sqrt{7\ap^2-2}}\beta \gamma\,,\\
j^-&=&
\Bigl( \am\sqrt{{1+\ap^2\over2}}\psi \d\Phi + \sqrt{2}bc\psi
\mp{\am\over\sqrt{2}}G_{\rm m} + {\am^2 - 2\over\sqrt{2}}\d \psi
\Bigr)e^{-\ap(\phi-\vst)}  \\ J^-&=&{} {1\over2\ap^2}
\Bigl( -\half(1+\ap^2)
\d\Phi\d\Phi + \half(3\ap - \am)\sqrt{1 + \ap^2}\d^2\Phi +
2\ap\sqrt{1+\ap^2}
b c \d\Phi \\ {}&{}&{}
- (1 - 3\ap^2) b\d c - (1 + \ap^2)\d b\,c + (1+\ap^2)
T_{\rm eff} \Bigr)e^{-2\ap(\phi-\vst)} \EA\label{supersl2}\EE where
useful combinations were introduced as
\BE\new\BA{rcl}
\d\Phi&=&
{\ap\over\sqrt{1 + \ap^2}}\Bigl( -(2\am + 3\ap )\d\phi  + ( \am + 3\ap )
        \d\vst  - \am\sqrt{7\ap^2-2}\beta\gamma \Bigr)\\
T_{\rm eff}&=&{1\over1+\ap^2}T_{\rm m}
 \pm{\ap\over1+\ap^2} G_{\rm m}\psi + {1 -
 2\ap^2\over2(1+\ap^2)}\d\psi\,\psi\,.\EA \label{Teff}\EE
One must be careful
to first evaluate in \req{supersl2} the normal-ordered product
${:}\d\Phi\d\Phi{:}$ and then,
${:}\,{:}\d\Phi\d\Phi{:}\,e^{-2\ap(\phi-\vst)}\,{:}$.
It can be checked that the currents \req{supersl2} satisfy the $osp(1|2)$
algebra:
\BE\new\BA{rclcrcl}
J^+(z) J^-(w)&=&{ k \over(z-w)^2} +{2J^0\over
z-w}, &{}&J^0(z)J^0(w)&=&{k/2\over(z-w)^2},\\ j^+(z) j^-(w) &=& {-2k
 \over(z-w)^2} - 2{J^0\over z-w}, &{}&J^0(z)J^\pm(w)&=&
\pm{J^\pm\over(z-w)}\\
 J^0(z) j^\pm(w) &=&\pm\half{j^\pm\over z-w}, &{}&J^\pm(z) j^\mp(w) &=& -
{j^\pm\over z-w}\,,\qquad
j^\pm(z)j^\pm(w)~{}={}~{}\mp2 {J^\pm\over z-w} \EA\EE where
now, in the supersymmetric setting, the level $k$ is introduced by
$k = (\am^2 - 3)/2$.

Next, the fermionic ($BC$) and bosonic ($\bbeta\ggamma$) ghosts are given
by
\BE\new\BA{rclcrcl} B&=& be^{-2\ap(\phi-\vst)}\,,&{}& C&=&
ce^{2\ap(\phi-\vst)}\,,\\ \bbeta&=& \beta
  e^{\sqrt{7\ap^2-2}(\phi-\vst)}\,,&{}& \ggamma&=& \gamma
  e^{-\sqrt{7\ap^2-2}(\phi-\vst)} \EA\label{BCBG}\EE

\smallskip

$\bullet$~Now let us consider
the ghosts from \req{BCBG}, along with the $osp(1|2)$
currents, as independent fields.  Note that, according to their explicit
constructions, $(B,C,\bbeta,\ggamma)$ are given dimensions $(1, 0, \half,
\half)$ by the \emt\ $\cTvee+T_\ast$.
Therefore, the appropriate \emt\
reads \BE T_{\rm ghosts}= -B\d C+\half\d\bbeta\,\ggamma-
\half\bbeta\d\ggamma
\EE Then, $T_{\rm ghosts}$ and the Sugawara \emt\ \BE T_{\rm Sug}= 2\ap^2(
J^0J^0 + \half J^+J^- + \half J^-J^+ + \fourth j^+j^- - \fourth j^-j^+)
+ \d
J^0 \label{sTsug}\EE describe the full system in terms of the `composite'
fields $(B, C, \bbeta, \ggamma, J^+, j^+, J^0, j^-, J^-)$.  Note that
$T_{\rm
Sug}$ assigns the currents $(J^+, j^+, J^0, j^-, J^-)$ dimensions $(0,
\half,
1, \threehalves, 2)$ respectively.
Evaluating $T_{\rm ghosts}+T_{\rm Sug}$
in terms of the `elementary' fields
$(T_{\rm m}, G_{\rm m}, b, c, \beta, \gamma, \d\phi, \psi, \d\vst)$,
one finds \BE T_{\rm Sug}+T_{\rm ghosts} =
\cTvee + T_\ast \label{equivalence}\EE This is some sort of the
`completeness
relation', showing that no degrees of freedom are lost when trading
supermatter coupled to supergravity and the additional $T_\ast$ piece
for an
$osp(1|2)$ algebra with the corresponding ghosts. This fact does not seem
obvious a priori, in particular considering that the matter theory is not
necessarily bosonised through a free superfield (nor are the $osp(1|2)$
currents free).

\subsection*{Applications to the KPZ Formulation}\lvm
The right-moving sector of the gauge-fixed theory is a conformal field
theory
consisting of the matter system, the $s\ell(2)_k$ current algebra
corresponding
to the gravitational degrees of freedom, and two ghost systems
corresponding
to the two gauge conditions \cite{[P-gr],[KPZ],[DJNW],[Itoh],[Ku],[Horv]}.
Choose
the matter sector to be a minimal model with \emt, $T'$, satisfying the
Virasoro algebra with \cc\ $d'=1-6{\an'}^2$ and define
$\ap'\am'=-1$\,, \ $\ap'+\am'=\an'$.

Primary states $\ket{r',s'}'$ of the matter sector are labelled by two
integers $r'$ and $s'$ and have dimensions
\BE\Delta'(r',s')=\fourth\Bigl(\ap'(r'-1)+\am'(s'-1)\Bigr)
\Bigl(\ap'(r'+1)+\am'(s'+1)\Bigr)\label{Deltaprime}\EE

In the ${sl}(2)_k$  sector,
a highest-weight spin-$j$ state $\ket j$
has twisted Sugawara dimension
\BE\tilde{\Delta^{\rm S}}(j)={j(j+1)\over k+2}-j\,.\label{Deltasug}\EE
Admissible $s\ell(2)_k$ representations are built upon highest-weight
states $\ket{j}$ whose spin is given by
$j=j_1-j_2(k+2)$ for some (half-) integers $j_1$ and $j_2$.

There are also two $bc$ ghost systems, $b\up2c\up2$ with spins
$2,-1$ and
$b\up0c\up0$ with dimensions $0,1$ respectively.
The total \emt\ is then \BE T^{\rm KPZ}=T'+\tilde{T^{\rm
S}}+t\up2+t\up0\,.\label{TKPZ}\EE The BRST current resulting from the
gauge-fixing is \BE\new\BA{rcl} \cJ^{\rm KPZ}_{\rm
BRST}(z)&=&c\up2\Bigl(T'+\tilde{T^{\rm S}}+\half t\up2+t\up0\Bigr)+c\up0J^+
\equiv
\cJ\up2(z)+\cJ\up0(z)\,.\EA \label{fullbrst}\EE For the BRST
charge $Q^{\rm KPZ}$ to be nilpotent, the
total \cc\ must vanish,
which determines $k_\pm=-2-{\alpha_\pm'}^2$\,.

Now, BRST-invariant states can be sought in the form
\BE\ket{r',s'}'\tensor\ket{j}_{s\ell(2)}\tensor
c_1\up2\ket0\up2\tensor\ket0\up0\,.  \label{astate}\EE
Vanishing of $Q^{\rm KPZ}$
on such a state determines $j$ in terms of $r'$ and $s'$.
There are two solutions related by
$\alpha_\pm'\leftrightarrow\alpha_\mp'$,
$r'\leftrightarrow s'$, and the one I choose to work with reads \BE
k=-2-{\am'}^2\,\qquad j(r',s')=\L\{\BA{l} {r'-1\over2}-(k+2){-s'-1\over2}\\
{}\\
{-r'-1\over2}-(k+2){s'-1\over2}\EA\R.\label{jlower}\EE

\medskip

$\bullet$~Consider now the KPZ theory in which the
$s\ell(2)\oplus u(1)_{b\up0c\up0}$ generators are realised in terms of the
topological ingredients
(a matter system $T$ dressed with a scalar $\phi$ and dimension-one
ghosts $b , c $) and the $\d\vst$ scalar,
The resulting theory is then precisely the DDK
theory of matter dressed with Liouville gravity, plus ghosts, tensored
with a
(`new') topological field theory. The latter is trivial when reduced by
the
action of appropriate screening charges, so that one obtains an explicit
construction of the DDK theory from the KPZ theory.
This works as follows.

The {\bf \emt} of the $s\ell(2)$ and dimension-zero ghost system
$\tilde{T^{\rm
S}}+t\up0 $ is given by \req{Tidentity},
with the
identification $B\equiv b\up0$, $C\equiv c\up0$.
Substituting this into the KPZ \emt\
given by \req{TKPZ} and using the construction \req{spin1} for the
topological \emt, eq.~\req{TKPZ} becomes \BE T^{{\rm
KPZ}}=T'+T-\d b\up2c\up2-2b\up2\d c\up2-\half(\d\phi)^2+{\an\over\st}\d^2\phi
-b \d c +T_\ast\,,
\label{basic}\EE
where, as in \req{Tidentity},
\BE T_\ast=\half(\d\vst)^2-{\an\over\sqrt{2}}\d^2\vst\,.\EE

Now notice that, by virtue of \req{jlower}, $d'=13+6/(k+2) +
6(k+2)$, whence $d+d'=26$ and therefore the first four terms in
\req{basic} are a good candidate to describe the DDK-type theory.
It actually
remains to compare the expression for the $s\ell(2)$ spin $j$, which
one gets from the KPZ theory,
with the expression for the same thing that follows from \req{ketket},
\req{jtrue}. It follows that {\it either\/} $(r'=-r,\,s'=s)$ {\it or\/}
$(r'=r,\,s'=-s)$, and therefore the dimensions in the matter and in the
matter$'$ sectors (i.e. those with \emt s $T$, $T'$)
add up to~1:
\BE\Delta'(r',s')+\Delta(r,s)=1\,,\label{addto1}\EE which is the DDK
prescription.  The two `matter' theories thus play the dual r\^oles of a
`true matter' and a `Liouville'.

Further, the \cc\ for the $\d\phi$-$\d\vst$-$b c $ sector vanishes and
this
sector is in
fact topological by itself. This is because its \emt\ is of the form of
the
one from \req{spin1}, but with its matter part $T$ replaced by $T_\ast$.
This $\ast$-matter has the same \cc\ $d=1-6\an^2$, which
fits
the Liouville \cc\ and the ghosts' dimensions to make up a topological
algebra according to the construction \req{spin1}.  The {\it only\/}
modification in the formulae for the new topological generators is the
replacement in \req{spin1} of\ $T$ with $T_\ast$.

It is the topological theory thus obtained that represents the
`difference'
between the KPZ and DDK formulations.
Equivalence of the KPZ and DDK
descriptions requires that the `extra' $\ast$-topological theory be empty.
However, in the way it has
emerged in \req{basic}, this theory seems to possess all the non-trivial
states given by a specialisation of the previous construction, namely by
eq.~\req{keth} in which $\ket{r,s}$ is now replaced by the free-field
realisation.  That is, the $\ast$-topological primary states are given by
\BE\ket{-p_{\rm M}(r,s)}_\ast\tensor\ket{p_{\rm
M}(r,s)}_{\rm L} \tensor\ket0_{bc}\,,\label{unwanted}\EE in which one
recognises \req{vphi}!

Recall, however, that
when using free-field constructions, the price to be paid for apparent
simplifications is the necessity of introducing screening (and/or
picture-changing) operators.
For example, the Wakimoto bosonisation of $s\ell(2)$ provides three
screening operators, of which two are bosonic and one fermionic.
This would
also have been the case with our representation \req{sl2} for the
$s\ell(2)$
curents, had the matter sector been bosonised through a scalar, via
$T\rightarrow\half(\d u)^2-{\an\over\st}\d^2 u$. Then, by a field
redefinition one would be able to map formulae~\req{sl2} into a
more
standard Wakimoto form \cite{[W],[Dko],[GMMOS]}, and the two standard
bosonic
screening operators from the Wakimoto representation would then be mapped
back into $e^{-\st\ap u}$ and $e^{-\st\am
u}$.
This shows that the screenings in our representation
\req{sl2} belong entirely to the matter sector and do not involve the
other
fields $\phi$, $\vst$, and $b c $ from \req{sl2}.

In general, $T$ in eqs.~\req{sl2}
can be the \emt\ of an arbitrary matter system.  Then, unless
$T$ is specifically chosen to be realised in terms of a Coulomb gas
representation, the corresponding screenings are not relevant to the
representation \req{sl2} of the $s\ell(2)_k$ algebra.  However, there
does exist
a sector in the theory that {\it is\/} bosonised through a scalar, and
therefore requires screening operators.  This is the $\ast$-topological
algebra given by the same construction~\req{spin1} in which one replaces
$T$
with $T_\ast$, the \emt\ for the scalar $\vst$.
The corresponding fermionic
screening current (cf.~\cite{[BLNW]}) has the form:
\BE\Sst=b e^{\am(\vst-\phi)/\st}\label{s}\EE This is completely
`OPE-isotropic', i.e. $\Sst(z)\Sst(w)=0$, and in addition satisfies the
OPE's
\BE\cJ(z)\Sst(w)=0\,,\qquad \cH(z)\Sst(w)=0\,.\label{HS}\EE
The nilpotent operator $\cQ\up\ast=\oint\Sst$
can be used as a BRST charge. There are, therefore, two BRST charges,
$\cQ$
and $ \cQ\up\ast$ and one can define different theories by demanding
that the
physical states be in the cohomology of $\cQ$ or of $ \cQ\up\ast$, or of
the
linear combination \ $Q^{\rm t.m.m.}=\cQ+\cQ\up\ast$, \ or one can demand
the
physical states to be in the double complex, i.e.  to be simultaneously in
the cohomology of both $\cQ$ and $ \cQ\up\ast$.  For example, whether
or not
a topological theory constructed by dressing matter with ghosts and
Liouville
is equivalent to a topological minimal model depends on the definition
of the
BRST operator, i.e. on whether or not $\cQ\up\ast$ is added to the BRST
charge \cite{[BLNW]}.

Consider the case in which physical states are simultaneously in the
cohomology of both $\cQ$ and $ \cQ\up\ast$, described in \cite{[BLNW]}
as a
`naive' case. This gives a much greater physical state space reduction.
This
makes sense in the current setup, because the $\ast$-topological sector is
only a part of a larger theory, and, according to \req{HS}, the action of
$\cQ\up\ast$ would preserve the $\cQ$-BRST invariance as well as
topological
$U(1)$ charges of topological primary fields. Moreover, with respect to
the
`full' \emt\ \req{Tidentity}, $\Sst$ is a screening current as well.

The $\ast$-topological algebra shares the BRST current $\cJ\equiv b $
with the topological algebra \req{spin1}.  A simple similarity
transformation
on the generators, \BE\cX_\ast\mapsto \biggl(\exp{\am\over\st}\oint
b c (\vst-\phi)\biggr)\cX_\ast \biggl(\exp-{\am\over\st}\oint
b c (\vst-\phi)\biggr) \label{rotation}\EE maps between the two
BRST currents, \ $\cJ\mapsto\Sst$, \ and changes the other generators
accordingly.  In this way, $\Sst$ becomes a BRST current of another
twisted
$N\!=\!2$ algebra.  All the states \req{unwanted}, corresponding to
operators
\req{vphi}, remain unchanged under this transformation, and therefore the
algebra resulting from \req{rotation} would, too, act on these states.
However, all these states are obviously BRST-trivial in this algebra, i.e.
with respect to $\cQ\up{\ast}=\oint\Sst$:  \BE
V_{r,s}=\{\cQ\up{\ast},c V_{r,s-1}\}\label{VQV}\EE

Therefore, in the $\ast$-sector, corresponding to the `coset' KPZ/DDK,
(i.e.,
roughly,
${\rm KPZ}={\rm DDK}\oplus \ast\mbox{-}{\rm topological}$), the states
\req{unwanted} are eliminated from the space of physical states by
`strengthening' the procedure of \cite{[BLNW]}, namely by choosing the
double
cohomology of $\cQ\up{\ast}$ and $\cQ$.  Note that the $\cQ\up{\ast}$
operator can obviously be lifted up to the `full' theory described by
\emt\
$\oT$ (it is essential that $[\oT(z),\cQ\up{\ast}]=0$).  In this sense
it can
be thought of as having been present in the theory from the start,
and the
corresponding reduction of the space of physical states makes a part of
the
definition of the theory.
With this definition adopted, one does indeed get a special gravitational
background of the KPZ theory on which it reduces to the DDK theory
tensored
with a {\it trivial\/} topological theory (consisting only of ground
states).
Once the operator $\cQ\up{\ast}$ is
`lifted up' to the whole KPZ theory, the operation performed in \req{VQV}
can
be written as $\{\cQ\up{\ast}, b\up0(J^+)^{-1}(\,\cdot\,)\}$, where the
current $\Sst$ itself becomes, formally,
\BE\Sst=c\up0(J^+)^{-k/2}\label{Sinvariant}\EE which might be interpreted
along the lines of ref.~\cite{[FM]}.

\medskip

$\bullet$~Now consider the {\bf $osp(1|2)$-supergravity}.
In the matter sector, one has an
$N\!=\!1$-supersymmetric matter${}''$-theory described by $T''_{\rm m}$
and
$G''_{\rm m}$ satisfying relations \req{matter} in which $d$ is replaced
by
$d'' = {15\over2} + 3\am^2 + 3\ap^2$.
Along with this matter system one
introduces \cite{[PZ]} a free Majorana--Weyl fermion $\chi$
with the \emt\ $T_\chi = \half\d\chi\,\chi$,
whence \cc\ $\half$.  Coupling matter${}''$ to
supergravity can be described then by an $osp(1|2)$ algebra and a set of
bosonic and fermionic ghosts: $b\up2c\up2$ as the usual reparametrisation
ghosts, $\beta\up{{3\over2}}\gamma\up{{3\over2}}$ as the super-ghosts,
and
also dimension-1 ghosts $b\up1c\up1$ and their superpartners
$\beta\up{{1\over2}}\gamma\up{{1\over2}}$, corresponding to the gauging of
$J^+$ and $j^+$ currents respectively.

Now, combine the  $osp(1|2)$ algebra with the $b\up1c\up1$ and
$\beta\up{{1\over2}}\gamma\up{{1\over2}}$ ghosts (to be identified with
$BC$
and $\bbeta\ggamma$ respectively) and express it in terms of the
topological
ingredients and the $\vst$-matter.  One will thus have the following field
content and the corresponding central charges:  \BE\new\BA{cccccccccc}
{\rm
matter}''&\chi&{\rm matter}&bc&\beta\gamma&\d\phi&\psi&\d\vst
&b\up2c\up2&\beta\up{{3\over2}}\gamma\up{{3\over2}}\\
\textstyle{15\over2}+3\ap^2+3\am^2&\textstyle{1\over2}&
                                   \textstyle{15\over2}-3\ap^2-3\am^2&
\textstyle-2&\textstyle-1&\textstyle-5+3\ap^2+3\am^2&
\textstyle{1\over2}&\textstyle7-3\ap^2-3\am^2
&\textstyle-26&\textstyle11 \EA\EE Here, $d''+d=15$, and the $b\up2c\up2$
and $\beta\up{{3\over2}}\gamma\up{{3\over2}}$ ghosts contribute
$-26+11=-15$.
One can therefore combine matter${}''$ + matter + $b\up2c\up2$ +
$\beta\up{{3\over2}}\gamma\up{{3\over2}}$ into a super-Distler--Kawai
sector,
in which the matter and matter${}''$ theories play the dual r\^oles of
`proper' matter
and Liouville (recall that each of the two matter sectors is
$N\!=\!1$-supersymmetric).

On top of that, another \cc-0 theory comprises \BE\BA{cccccc} bc& \d\phi&
\d\vst&\chi&\beta\gamma&\psi\\
-2&(-5+3\ap^2+3\am^2)&(7-3\ap^2-3\am^2)&{1\over2}&-1&{1\over2} \EA\EE
Here,
$\d\vst$ and $\chi$ can be combined into a supersymmetric $\ast$-matter,
which is then dressed with the other fields into a twisted $N\!=\!3$
algebra
as above, the difference from \req{GGG} and \req{KKK} being that the
$T_{\rm m}$ and $G_{\rm m}$ generators are now given in terms of the
free fields:
\BE{T_{\rm m}}_\ast =\half\d\vst\d\vst + \half(\ap +
\am)\d^2\vst + \half\d\chi\,\chi\,,\qquad
{G_{\rm m}}_\ast = \d\vst\,\chi  + (\ap + \am) \d\chi \EE

A characteristic difference from the bosonic construction is the fact that
the $\d\vst$ theory is supersymmetrised by a fermion, $\chi$, that comes
from
the $osp(1|2)$ gravity.  This `corrects' the fact that in the super-matter
sector, the fermionic part brings in a $\half$, whereas $\d\vst$ is only a
free boson, with a $\half$ thus missing from the central charge.  The
$\d\vst$ field does actually get
its superpartner from the `extra ghost' of the
$osp(1|2)$ gravity. Reversing the argument, i.e. constructing the
$osp(1|2)$
currents and the associated ghosts from the super-Distler--Kawai fields
and
the $\ast$-$N\!=\!3$ algebra, we thus see that this extra ghost is nothing
but a survived superpartner to the $\d\vst$ matter.

\smallskip

We have seen, among other things, that, just as a large class of string
theories have a twisted $N\!=\!2$ superconformal algebra, a certain class of
conformal field theories have a hidden $sl(2)$ Ka\v{c}--Moody
algebra.

\smallskip
\noi
Acknowledgements. It is a pleasure to thank
D.~L\"ust and H.~Dorn and the organisers of
the Symposium on the Theory of Elementary Particles in Wendisch-Rietz.
The
research reviewed in this talk
was made possible in part by Grant \#MQM000
from the International Science Foundation
and by an RFFI grant.

\end{document}